\documentclass{PoS}
\usepackage[T1]{fontenc}
\usepackage{graphicx}
\usepackage{subfigure}
\usepackage{amsmath}

\title{$B\to\pi\ell\nu$ and $B\to\pi\ell^+\ell^-$ semileptonic form factors from unquenched lattice QCD}

\ShortTitle{$B\to\pi$ form factors from lattice QCD}

\author{\speaker{Daping~Du} $^{a}$\thanks{Present address: Department of Physics, Syracuse University, Syracuse, New York, USA}, 
Jon A. Bailey$^b$, A. Bazavov$^c$\thanks{Present address: Department of Physics and Astronomy, University of Iowa, Iowa City, IA 52245}, C. Bernard$^d$, A. X. El-Khadra$^{a,e}$, Steven~Gottlieb$^f$, R. D. Jain$^a$, A. S. Kronfeld$^e$, 
J. Laiho$^{g\dagger}$, Yuzhi Liu$^{h}$\thanks{Present address: Department of Physics, University of Colorado, Boulder, CO 80309, USA}, P.~B.~Mackenzie$^e$, Y.~Meurice$^h$, R. S. Van de Water$^e$, Ran Zhou$^{e,f}$ \\
\llap{$^a$} Physics Department, University of Illinois, Urbana, Illinois, USA\\
\llap{$^b$} Department of Physics and Astronomy, Seoul National University, Seoul, South Korea \\
\llap{$^c$} Physics Department, Brookhaven National Laboratory \thanks{Brookhaven National Lab is operated by Brookhaven Science Associates, LLC, under Contract No. DE-AC02-98CH10886 with the U.S. De-partment of Energy. }, Upton, New York, USA \\
\llap{$^d$} Department of Physics, Washington University, St. Louis, Missouri, USA \\
\llap{$^e$} Fermi National Accelerator Laboratory \thanks{Fermilab is operated by Fermi Research Alliance, LLC, under Contract No. DE-AC02-07CH11359 with the U.S. Department of Energy.}, Batavia, Illinois, USA\\
\llap{$^f$} Department of Physics, Indiana University, Bloomington, Indiana, USA\\
\llap{$^g$} SUPA, Department of Physics and Astronomy, University of Glasgow, Glasgow, G12~8QQ, United 
Kingdom \\
\llap{$^h$} Department of Physics and Astronomy, University of Iowa, Iowa City, Iowa, USA\\
E-mail: \email{ddu@illinois.edu} } 
\author{(Fermilab Lattice and MILC Collaborations)  }

\abstract{We update the lattice calculation of the $B\to\pi$ semileptonic form factors, which have important applications to the CKM matrix element $|V_{ub}|$ and the $B\to\pi\ell^+\ell^-$ rare decay. We use MILC asqtad ensembles with $N_f=2+1$ sea quarks and over a range of lattice spacings $a \approx 0.045$--$0.12$~fm. We perform a combined chiral and continuum extrapolation of our lattice data using SU(2) staggered chiral perturbation theory in the hard pion limit. To extend the results for the form factors to the full kinematic range, we take a functional approach to parameterize the form factors using the Bourrely-Caprini-Lellouch formalism in a model-independent way. Our analysis is still blinded with an unknown off-set factor which will be disclosed when we present the final results.  }

\FullConference{31st International Symposium on Lattice Field Theory - LATTICE 2013\\
		July 29 - August 3, 2013\\
		Mainz, Germany}

\begin{document}

\section{ Introduction}
\label{sec:1}

The Cabibbo-Kobayashi-Maskawa (CKM) matrix element $|V_{ub}|$ is an important Standard Model (SM) parameter that can be determined through the experimentally measured differential decay rate of the exclusive $B\to \pi \ell\nu$ decay
\begin{eqnarray} \label{eqn:partial_rate}
\frac{d\Gamma(B\to\pi\ell\nu)}{dq^2} &=& \frac{G_F^2|V_{ub}|^2}{24\pi^3}|{\bf p}_\pi|^3 |f_+(q^2)|^2,
\end{eqnarray}
if the form factor $f_+$ is known from theory. The form factors $f_+$ and $f_0$, which parametrize the hadronic matrix element $\langle \pi |\mathcal{V}^\mu |B\rangle$, encode the non-perturbative QCD effects in the kinematic dependence and can be reliably calculated using lattice QCD \cite{ElKhadra:2001rv, Dalgic:2006dt, Bailey:2008wp}. The unitarity test of the CKM matrix requires that the off-diagonal elements such as $|V_{ub}|$ be known to high precision. There is a long-standing tension between the value of $|V_{ub}|$ determined from exclusive and inclusive methods. To address this challenge, it is important to improve upon the existing lattice calculations as well as the experimental measurements~\cite{Lees:2012vv}. In particular, the quantity $f_+$ from lattice QCD has not been updated (in the peer-reviewed literature) since 2008 \cite{Bailey:2008wp}. Recently, several efforts (including this one) from different lattice collaborations \cite{Kawanai:2012id} have been aiming to improve the determination of $f_+$ with new data (better statistics, smaller lattice spacings and smaller light quark masses) and improved theoretical methods. Another topic that motivates this calculation is the rare $B\to \pi \ell^+\ell^-$ decay, which is loop-suppressed in the SM, and therefore sensitive to new physics. The low-energy effective operators that contribute to this process in the SM are the flavor-changing vector and tensor currents. Lattice calculations of vector and tensor form factors for this process are timely, since first experimental measurements have already appeared \cite{LHCb:2012de}. Thus, the calculation of $f_T$ is also a focus of this analysis. 

\section{Lattice calculation}
\label{sec:2}

Our calculation is based on a subset of the MILC (2+1)-flavor asqtad ensembles \cite{Bazavov:2009bb} that have large numbers of configurations (ranging from 593 to 2259). We use 12 ensembles at four different lattice spacings (roughly $0.12$, $0.09$, $0.06$ and $0.045$ fm) with the light quark over strange quark mass ratio as low as 0.05. The details of these ensembles are summarized in Table \ref{tab:ensembles}. The light asqtad valence quarks use the same masses as in the sea, while
the $b$ quark uses the Sheikholeslami-Wohlert clover action with the Fermilab interpretation \cite{ElKhadra:1996mp}. 

\begin{table}[h]
\centering
\caption{MILC asqtad ensembles and their simulation parameters used in this analysis. The columns are, from left to right, lattice spacing $a$ in fm, the light/strange quark mass ratio $am_l/am_s$, lattice size space$\times$time, number of configurations $N_{\text{cfg}}$, root-mean-squared (RMS) pion mass $M_\pi$, the Goldstone pion mass $M_\pi$, $M_\pi L$ ($L$ is the size of space), the source-sink separations $t_{sink}$ and the $b$ quark mass (hopping) parameter $\kappa_b$.\label{tab:ensembles} }
\begin{tabular}{ccccccccc}
\hline\hline
$a$(fm)     & $am_l/am_s$   & Size 				 & $N_{\text{cfg}}$	& $M^\text{RMS}_\pi$(MeV) &$M_\pi$(MeV) & $M_\pi L$ &  $t_\text{sink}$  & $\kappa_b$ \\
\hline
$\sim$0.12	&	0.2 		&	 $20^3\times 64$ &	 2259  			& 	532			&389			& 	4.5		&	18,19			& 0.0901     \\
			&	0.14 	    &    $20^3\times 64$ &   2110 			&	488			&327			&  	3.8		&	18,19			&0.0901    \\
			&	0.1 	    &    $24^3\times 64$ &   2099 			&	456			&277			&  	3.8		&	18,19			&0.0901    \\
\hline                       
$\sim$0.09	&	0.2    		&    $28^3\times 96$ &   1931 			&	413			&354			&  	4.1		&	25,26			&0.0979    \\
			&	0.15   		&    $32^3\times 96$ &   984  			&	374			&307			&  	4.1		&	25,26			&0.0997    \\
			&	0.1   		&    $40^3\times 96$ &   1015 			&	329			&249			&  	4.2		&	25,26			&0.0976    \\
			&	0.05   		&    $64^3\times 96$ &   791  			&	277			&177			&  	4.8		&	25,26			&0.0976    \\
\hline                       
$\sim$0.06	&	0.4    		&    $48^3\times 144$&   593  			& 	466			&450			&  	6.3		&	36,37			&0.1048   \\
			&	0.2    		&    $48^3\times 144$&   673  			& 	340			&316			&  	4.5		&	36,37			&0.1052   \\
			&	0.14    	&    $56^3\times 144$&   801  			& 	291			&264			&  	4.4		&	36,37			&0.1052   \\
			&	0.1    		&    $64^3\times 144$&   827  			& 	255			&224			&  	4.3		&	36,37			&0.1052   \\
\hline                       
$\sim$0.045	&	0.2    		&    $64^3\times 192$&   801  			& 	331			&324			&  	4.6		&	48,49			&0.1143   \\
\hline\hline
\end{tabular}
\end{table}

\begin{figure}[h]
\centering
\hfill
\subfigure{\includegraphics[width=0.45\textwidth]{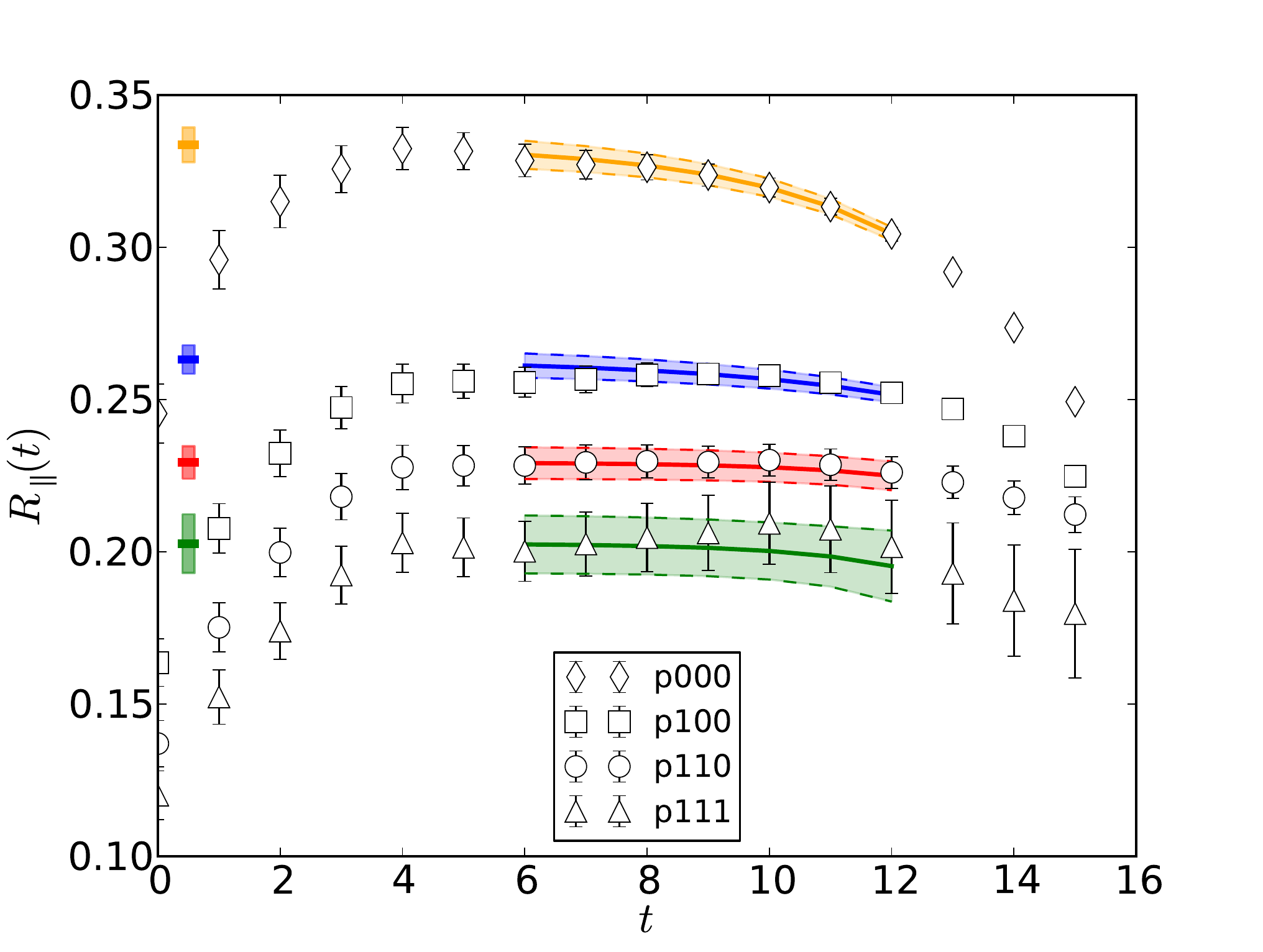} }
\hfill
\subfigure{\includegraphics[width=0.45\textwidth]{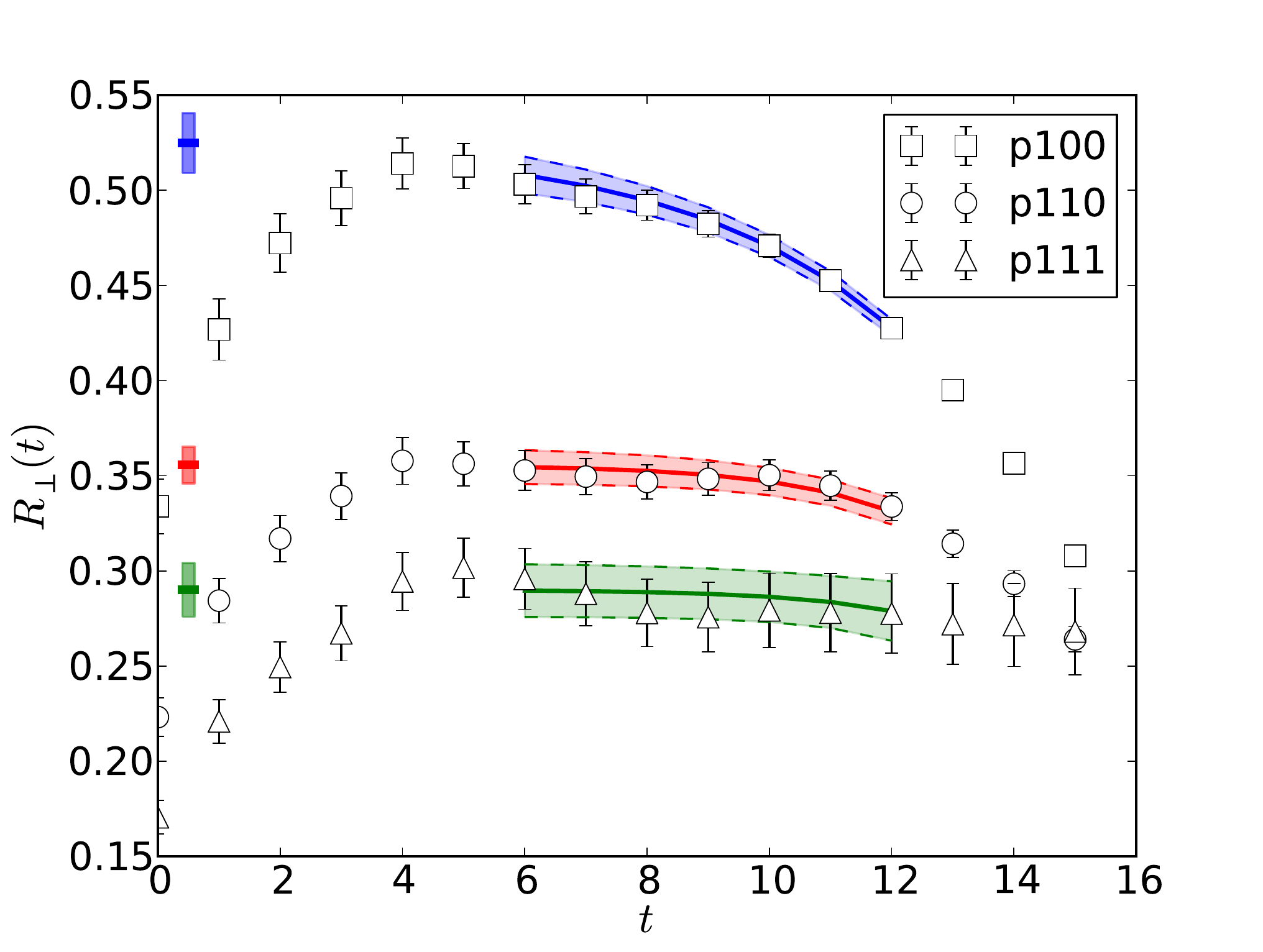}}
\hfill
\caption{Plots of the averaged ratio $R_\parallel (t)$ (left) and $R_\perp (t)$ (right) and their fit results for the ensemble ($a\approx 0.12$~fm, 0.1~$m_s$). The data points with error bars are the ratios constructed from two-point and three-point correlation functions with various momenta. The colored bands show the best fit and error for each momentum. The
horizontal extent indicates the fit range. The fit results (constants in ratio $R_{\parallel (\perp)}$ and their errors) are marked as the color bars on the left close to the axis. }
\label{fig:ratios}
\end{figure}

The relevant operators in our calculation are the vector current $V^\mu_{\text{lat}}=\bar{q}\gamma^\mu b $ and tensor current 
$T_{\text{lat}}^{\mu\nu} = i\bar{q}\sigma^{\mu\nu}b$, which are related to the continuum currents by renormalization factors such that $\langle \pi |\Gamma_\text{cont}|B\rangle = Z^{hl}_\Gamma \langle \pi | \Gamma_\text{lat} | B\rangle $ where $\Gamma = V^\mu$ or $T^{\mu\nu}$. We determine $Z^{hl}_\Gamma$ through the relation $Z^{hl}_\Gamma = \rho_\Gamma^{hl} \sqrt{Z_\Gamma^{hh}Z_\Gamma^{ll}}$ \cite{ElKhadra:2001rv} where $Z_\Gamma^{hl}$ is dominated by the non-perturbatively calculated factors $Z_\Gamma^{hh}, Z_\Gamma^{ll}$. The flavor-changing part of the renormalization is captured by $\rho_\Gamma^{hl}$ which is determined using lattice perturbation theory. Note that our whole analysis is currently blinded by a constant factor multiplying $\rho_\Gamma^{hl}$. We parameterize the vector-current matrix elements in terms of the form factors $f_\parallel$ and $f_\perp$, 
\begin{eqnarray}
\langle \pi | V_\text{lat}^\mu | B\rangle = \sqrt{ 2 M_B} \left [ v^\mu f_\parallel(E_\pi) + p_\perp^\mu f_\perp(E_\pi) \right ], 
\end{eqnarray}
where $v^\mu=p_B^\mu/M_B$ and $p_\perp^\mu = p_\pi^\mu - (p_\pi\cdot v)v^\mu$. $f_{\parallel, \perp}$ can easily be converted to the phenomenologically relevant $f_{+,0}$ \cite{ElKhadra:2001rv}. 
We measure the two-point and three-point correlation functions, 
explicitly given by
\begin{eqnarray}
C_{2pt}^P(t; {\bf p}) &=& \sum_{\bf x} e^{i {\bf p}\cdot {\bf x}} \langle \mathcal{O}_P (0, {\bf 0}) \mathcal{O}^\dagger_P(t,{\bf x})\rangle, \nonumber \\
C_{3pt}^{\Gamma} (t, t_{sink}; {\bf p}) &=& \sum_{\bf x, y} e^{i {\bf p}\cdot {\bf y}} \langle \mathcal{O}_\pi (0, {\bf 0}) \Gamma_\text{lat}(t,{\bf y})\mathcal{O}^\dagger_B(t_{sink},{\bf x})\rangle ,
\end{eqnarray}
where $P=\pi, B$ and $\Gamma_\text{lat} = V^{\mu}_\text{lat}, T^{\mu\nu}_\text{lat}$. The source-sink separation $t_{sink}$, which corresponds to roughly the same physical separation at each lattice spacing, has been optimized to maximize the signal/noise ratio. However we vary the source-sink separations by one unit, $i.e.$, using $t_{sink}$ and $t_{sink}+1$, to control the staggered oscillating state contributions in the correlator fits. The data with different source-sink separations is necessary to suppress the oscillating states with the wrong parity, which are the artifacts due to staggered light quark action used. For that purpose, we construct the average of the correlation functions introduced in Eqs.~(37) and (38) of Ref.~\cite{Bailey:2008wp}, denoted by $\overline{C}$. Finally, we extract the form factors by constructing the ratios \cite{Bailey:2008wp}
\begin{eqnarray} \label{eq:ratios}
R_\Gamma (t) = \frac{\overline{C}_{3pt}^\Gamma (t, t_{sink})} {\sqrt{\overline{C}_{2pt}^\pi(t) \overline{C}^B_{2pt}(t_{sink}-t)}} \sqrt{ \frac{2 E_\pi}{e^{-E_\pi^{(0)} t}e^{-M_B^{(0)}(t_{sink}-t)}}},
\end{eqnarray}
where $\Gamma = \parallel, \perp, T$, denotes the three-point functions $C_{3pt}^{V^0}, C_{3pt}^{V^i}, C_{3pt}^{T^{0i}}$, respectively. 
The ratios defined in Eq. (\ref{eq:ratios}) have the advantage that the relevant wave function factors cancel, but the tradeoff is that we need an additional factor (the whole square root on the right) to suppress the time dependence by using the ground state energies $E_\pi^{(0)}$ and $M_B^{(0)}$. If the ground states are overwhelmingly dominant, then the ratios $R_\Gamma$ are independent of $t$ and give $f_\Gamma^{\rm lat}$ up to constant factors. However, with our statistical errors the excited state contributions in the $B\to\pi$ data turn out to be significant and have to be included in the fits to avoid large systematic errors. We find that the first excited state of the $B$ meson accounts for most of the excited state contribution and the fit yields consistent results compared to fits that include more excited states of both the pion and $B$ meson. Thus, we fit the ratios to the following ansatz
\begin{eqnarray}
R_\Gamma (t)/k_\Gamma =  f_\Gamma^\text{lat} \, \left [ 1 + \mathcal{A}_\Gamma \, e^{-\Delta M_{B} (t_{sink}-t)} \right ], 
\end{eqnarray}
where $\mathcal{A}_\Gamma$ are fit parameters, $\Delta M_B$ is the lowest energy splitting of the $B$ meson and the factors on the left are $k_\parallel = 1, k_\perp=|p_\pi^i|, k_T = (\sqrt{2M_B}|p_\pi^i|)/(M_B+M_\pi)$. Figure \ref{fig:ratios} shows examples of the fits of ratios $R_\parallel$ and $R_\perp$ ($R_T$ follows rather similarly).

\section{Chiral and continuum extrapolation}
\label{sec:3}

\begin{figure} [hd]
\includegraphics[ width=0.98\textwidth]{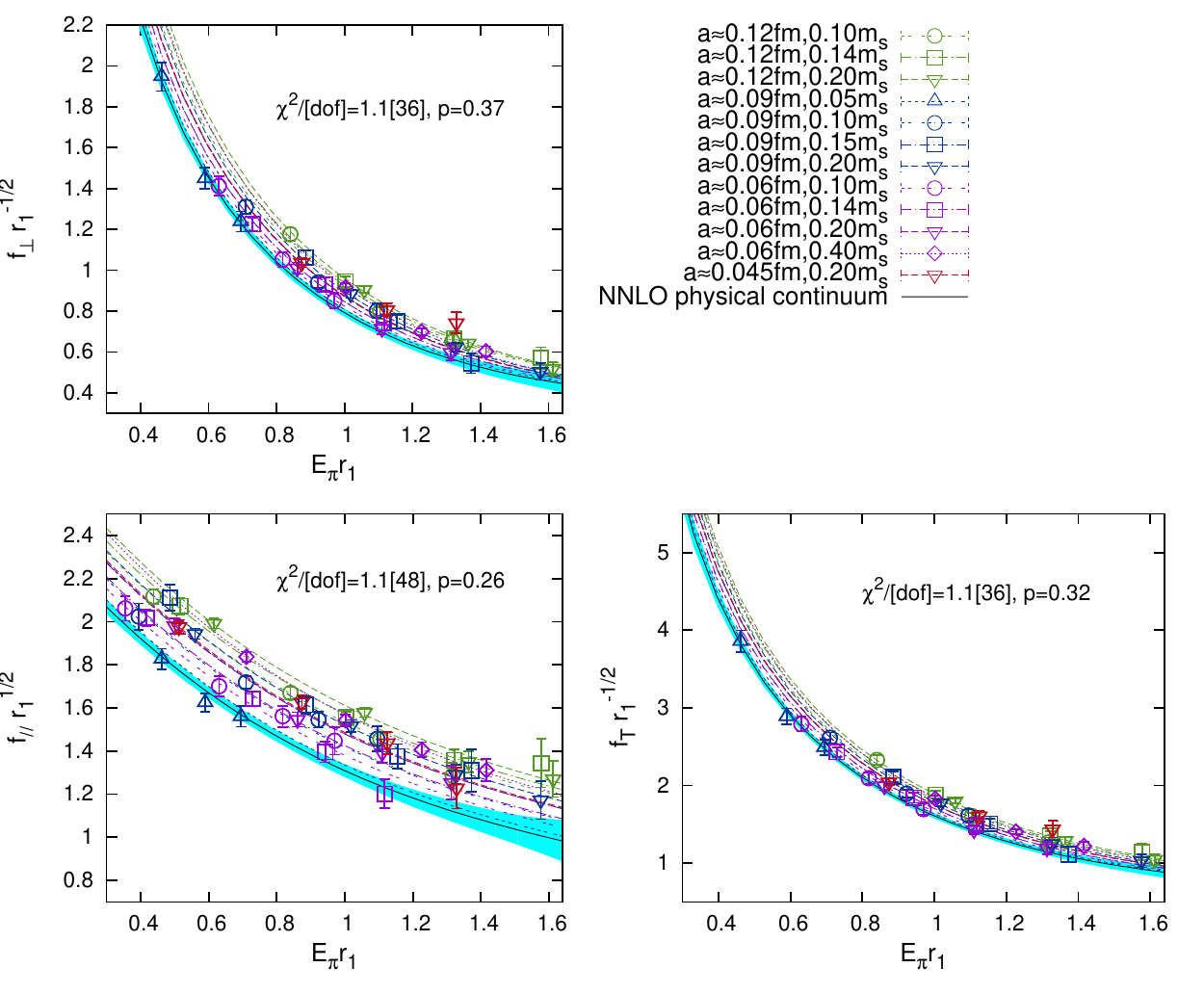} 
\hfill
\caption{ Chiral-continuum fit results for the form factors $f^\text{lat}_\perp$ (upper left), $f^\text{lat}_\parallel$ (lower left) and $f^\text{lat}_T$ (lower right) in $r_1$ units. We plot our form factor data using color to indicate the lattice spacing and
shape to indicate the ratio $m_l/m_s$, as detailed in the legend. The black solid lines in shaded bands are the $\chi$PT-continuum extrapolated curves with their statistical errors.   \label{fig:chiral}}
\end{figure}

Our chiral and continuum extrapolation is based on heavy meson staggered chiral perturbation theory (HMS$\chi$PT) \cite{Aubin:2007mc}, but with some modifications. The HMS$\chi$PT is derived with the assumption that the external pion and the pions in the loop should be soft ($E_\pi \sim M_\pi$); however, the pions with non-zero momenta in the simulation are mostly too energetic. As a result, the HMS$\chi$PT provides a poor description of our data for $f_\parallel^\text{lat}$. Thus, we adapt the hard-pion $\chi$PT \cite{Bijnens:2010ws} by incorporating the taste-breaking discretization effects from staggered fermions. In addition, it was argued \cite{Becirevic:2007dg} that the SU(2) $\chi$PT is more justified than the $SU(3)$ $\chi$PT for lattice data. It turns out that the next to leading order (NLO) HMS$\chi$PT in the hard pion and SU(2) limit gives reasonable fits to our lattice form factors $f_{\perp, \parallel, T}$, and we find that the systematic error due to higher-order chiral corrections is largely captured by the statistical errors in the fits that include the NNLO (next to NLO) analytic terms. Thus, we use the NNLO (analytic terms only) hard-pion and SU(2) HMS$\chi$PT fits as our standard fits. The results are shown in Fig. \ref{fig:chiral}. 

\section{ A new functional approach to the $z$ expansion}

\begin{figure}[h]
\centering
\includegraphics[ width=0.73\textwidth]{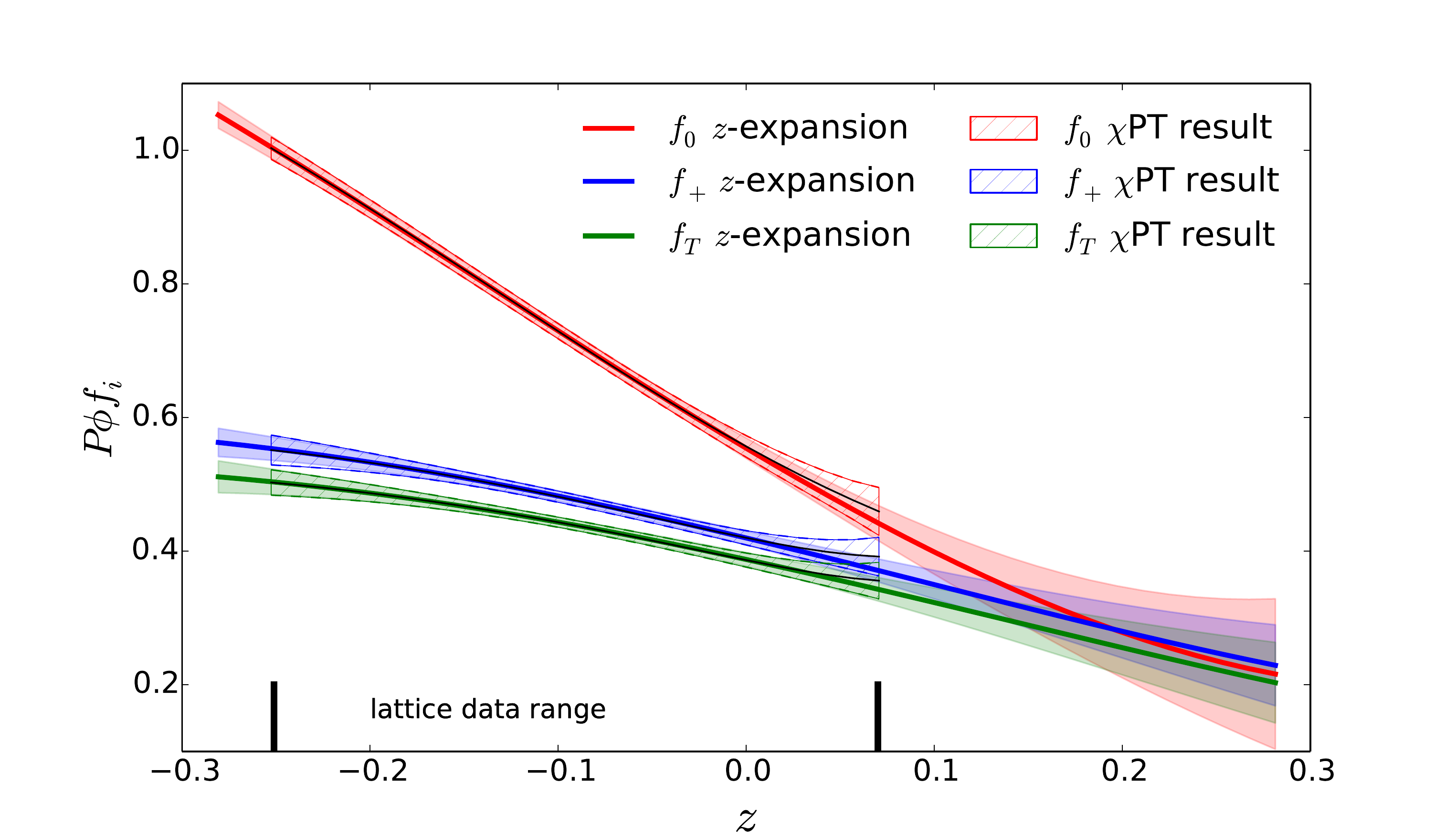} 
\hfill
\caption{The $\chi$PT-continuum fit results of $f_{0,+,T}$ (black solid lines with hatched error bands ) are extrapolated to the full kinematic range (colored solid lines in shaded error bands) using the functional $z$ expansion method. The form factors $f_{0}$, $f_+$, $f_T$ (from top to bottom) are plotted with the pole structure removed by the factors $P\phi$ in front. \label{fig:zfit} }
\end{figure}

To extend the form factors $f_{+,0,T}$ (constructed from the lattice form factors $f_{\parallel, \perp, T}$ in the physical and continuum limit with the appropriate renormalization factors) to the full kinematic range, we use the $z$-expansion method in the Bourrely-Caprini-Lellouch (BCL) formalism \cite{Bourrely:2008za}. Explicitly,
\begin{eqnarray} \label{eq:BCL}
f_{+,T}= \frac{1}{1-q^2/M_{B^*}^2} \sum_{n=0}^{N-1} b^{+,T}_n \left ( z^n-(-1)^{N-n}\frac{n}{N} z^{N} \right),\;\;\;
f_0 = \sum_{n=0}^{N-1} b^0_n z^n,
\end{eqnarray}
where $N$ is the truncation order and the expansion of $f_0$ is simple due to the fact that it has no poles below the pair-production threshold. The reparameterization is normally done by taking synthetic data at several kinematic points from the $\chi$PT-continuum fit results $f^{\chi\text{PT}}_i$ ($i=+,0,T$) and fitting them to Eq. (\ref{eq:BCL}) using the variable $z$. In this analysis, instead of taking synthetic points, we consider the independent function forms in $f^{\chi\text{PT}}_i(z)$. The correlation in $f^{\chi\text{PT}}_i(z)$ is represented by a kernel function $K_i(z,z') = E[ \delta f^{\chi\text{PT}}_i(z) \delta f^{\chi\text{PT}}_i(z') ] $ where $\delta f^{\chi\text{PT}}_i(z)$ is the fluctuation of the function at $z$, and $E[\cdot]$ denotes the statistical expectation. Since there are only a few independent functional forms in $f^{\chi\text{PT}}_i(z)$, the Mercer kernel $K_i(z,z')$ has a finite orthonormal representation \cite{mercer:1909}, based on which we can construct the $z$-expansions. Explicitly, we determine the coefficients for the $z$-expansions by minimizing
\begin{eqnarray} \label{eq:funcz}
\int dz \int dz' \; [ f^{\chi\text{PT}}_i(z) - f_i(z) ] K_i^{-1}(z,z') [ f^{\chi\text{PT}}_i(z') - f_i(z') ],
\end{eqnarray}
where $f_i(z)$ are given in Eq. (\ref{eq:BCL}). Equation (\ref{eq:funcz}) is a functional analog of the common chi-squared statistic for discrete data. The range of integration covers that of the lattice data, and reasonable variations of the range have negligible effects. The benefit of this functional approach is that the extrapolation is very robust against the unphysical behaviors of the lattice form factors in the large-$E_\pi$ region where the chiral expansion fails. The expansion coefficients $b_n$'s for $f_{+,T}$ in Eq. (\ref{eq:BCL}) are constrained by analyticity \cite{Becher:2005bg} and the pole-dominant feature of the form factors, while those for $f_0$ are constrained by the weaker unitarity condition \cite{Boyd:1997qw}. We vary the order at which the $z$-expansion is truncated and find that the results (central values and errors) are stable for $N \geq 4$ and therefore truncate the series at $N=4$.  The result for the three form factors is shown in Fig.~\ref{fig:zfit}. The form factors $f_+$ and $f_0$ in Fig.~\ref{fig:zfit} are obtained through separate $z$-expansions; however, the kinematic condition $f_+(q^2=0) = f_0(q^2=0)$ is satisfied naturally. We find a high degree of correlation between $f_+$ and $f_T$, which is expected because they approach the same heavy quark limit.

\section{Discussion and outlook}
\label{sec:4}

We are currently finalizing our error budgets for the form factors $f_+$, $f_0$, and $f_T$.  We anticipate that our largest uncertainties will be from statistics and the $\chi$PT-continuum extrapolation. Our next step is to consider all the possible sources of systematic uncertainties and present our results with full error budgets. Once our analysis is final, we will unblind our form factor results and discuss their implications for SM phenomenology.

\section*{Acknowledgements}

This work was supported by the U.S. Department of Energy, National Science Foundation and Universities Research Association. Computation for this work was done at the Argonne Leadership Computing Facility (ALCF), the
National Center for Atmospheric Research (UCAR), the National Center for Supercomputing Re-
sources (NCSA), the National Energy Resources Supercomputing Center (NERSC), the National
Institute for Computational Sciences (NICS), the Texas Advanced Computing Center (TACC), and
the USQCD facilities at Fermilab, under grants from the NSF and DOE.


\end{document}